# Achieving QoS for TCP traffic in Satellite Networks with Differentiated Services*

Arjan Durresi[1], Sastri Kota[2], Mukul Goyal[1], Raj Jain[3], Venkata Bharani[1]

[1]Department of Computer and Information Science, The Ohio State University,
2015 Neil Ave, Columbus, OH 43210-1277, USA
Tel: 614-688-5610, Fax: 614-292-2911, Email: {durresi, mukul, bharani}@cis.ohio-state.edu

[2]Lockheed Martin
60 East Tasman Avenue MS: C2135
San Jose CA 95134, USA
Tel: 408-456-6300/408- 473-5782, Email: sastri.kota@lmco.com

[3]Nayna Networks, Inc.
157 Topaz St., Milpitas, CA 95035, USA
Tel: 408-956-8000X309, Fax: 408-956-8730, Email: raj@nayna.com

**ABSTRACT**

Satellite networks play an indispensable role in providing global Internet access and electronic connectivity. To achieve such a global communications, provisioning of quality of service (QoS) within the advanced satellite systems is the main requirement. One of the key mechanisms of implementing the quality of service is traffic management. Traffic management becomes a crucial factor in the case of satellite network because of the limited availability of their resources. Currently, Internet Protocol (IP) only has minimal traffic management capabilities and provides best effort services. In this paper, we presented a broadband satellite network QoS model and simulated performance results. In particular, we discussed the TCP flow aggregates performance for their good behavior in the presence of competing UDP flow aggregates in the same assured forwarding. We identified several factors that affect the performance in the mixed environments and quantified their effects using a full factorial design of experiment methodology.

Correspondence address: Dr. Arjan Durresi,
Department of Computer and Information Science,
The Ohio State University, 2015 Neil Ave, Columbus, OH 43210, USA
Tel: 614-688-5610, Fax: 614-292-2911, Email: durresi@cis.ohio-state.edu

* This work was supporeted in part by grants from OAI, Cleveland, Ohio and NASA Glenn Research Center, Cleveland

## 1. INTRODUCTION

The increasing worldwide demand for more bandwidth and Internet access is creating new opportunities for the deployment of global next generation satellite networks. Today it is clear that satellite networks will be a significant player in the digital revolution, and will specially benefit from on-board digital processing and switching, as well as other such technological advances as emerging digital compression, narrow spot beams for frequency reuse, digital intersatellite links, advanced link access methods and multicast technologies. Many new satellite communication systems have been planned and are under development including at Ka, Q/V-bands [7]. Some of the key design issues for satellite networks include efficient resource management schemes and QoS architectures.

However, satellite systems have several inherent constraints. The resources of the satellite communication network, especially the satellite and the Earth station, are expensive and typically have low redundancy; these must be robust and be used efficiently. The large delays in geostationary Earth orbit (GEO) systems and delay variations in low Earth orbit (LEO) systems affect both real-time and non-real-time applications. In an acknowledgement and time-out-based congestion control mechanism (like TCP), performance is inherently related to the delay-bandwidth product of the connection. Moreover, TCP round-trip time (RTT) measurements are sensitive to delay variations that may cause false timeouts and retransmissions. As a result, the congestion control issues for broadband satellite networks are somewhat different from those of low-latency terrestrial networks. Both interoperability issues as well as performance issues need to be addressed before a transport-layer protocol like TCP can satisfactorily work over long-latency satellite IP ATM networks.

There has been an increased interest in developing Differentiated Services (DS) architecture for provisioning IP QoS over satellite networks. DS aims to provide scalable service differentiation in the Internet that can be used to permit differentiated pricing of Internet service [1]. This differentiation may either be quantitative or relative. DS is scalable as traffic classification and conditioning is performed only at network boundary nodes. The service to be received by a traffic is marked as a code point in the DS field in the IPv4 or IPv6 header. The DS code point in the header of an IP packet is used to determine the Per-Hop Behavior (PHB), i.e. the forwarding treatment it will receive at a network node. Currently, formal specification is available for two PHBs - Assured Forwarding [4] and Expedited Forwarding [5]. In Expedited Forwarding, a transit node uses policing and shaping mechanisms to ensure that the maximum arrival rate of a traffic aggregate is less than its minimum departure rate. At each transit node, the minimum departure rate of a traffic aggregate should be configurable and independent of other traffic at the node. Such a per-hop behavior results in minimum delay and jitter and can be used to provide an end-to-end `Virtual Leased Line' type of service.

In Assured Forwarding (AF), IP packets are classified as belonging to one of four traffic classes. IP packets assigned to different traffic classes are forwarded independent of each other. Each traffic class is assigned a minimum configurable amount of resources (link bandwidth and buffer space). Resources not being currently used by another PHB or an AF traffic class can optionally be used by remaining classes. Within a traffic class, a packet is assigned one of three levels of drop precedence (green, yellow, red). In case of congestion, an AF-compliant DS node drops low precedence (red) packets in preference to higher precedence (green, yellow) packets.

In this paper, we describe a wide range of simulations, varying several factors to identify the significant ones influencing fair allocation of excess satellite network resources among congestion sensitive and insensitive flows. The factors that we studied in Section 2 include *a)* number of drop precedence required (one, two, or three), *b)* percentage of reserved (highest drop precedence) traffic, *c)* buffer management (Tail drop or Random Early Drop with different parameters), and *d)* traffic types (TCP aggregates, UDP aggregates). Section 3 describes the simulation configuration and parameters and experimental design techniques. Section 4 describes Analysis Of Variation (ANOVA) technique. Simulation results for TCP and UDP, for reserve rate utilization and fairness are also given. Section 5 summarizes the study's conclusions.

## 2. QOS FRAME WORK

The key factors that affect the satellite network performance are those relating to bandwidth management, buffer management, traffic types and their treatment, and network configuration. Band width management relates to the algorithms and parameters that affect service (PHB) given to a particular aggregate. In particular, the number of drop precedence (one, two, or three) and the level of reserved traffic were identified as the key factors in this analysis.

Buffer management relates to the method of selecting packets to be dropped when the buffers are full. Two commonly used methods are tail drop and random early drop (RED). Several variations of RED are possible in case of multiple drop precedence. These variations are described in Section 3.

Two traffic types that we considered are TCP and UDP aggregates. TCP and UDP were separated out because of their different response to packet losses. In particular, we were concerned that if excess TCP and excess UDP were both given the same treatment, TCP flows will reduce their rates on packet drops while UDP flows will not change and get the entire excess bandwidth. The analysis shows that this is in fact the case and that it is important to give a better treatment to excess TCP than excess UDP.

In this paper, we used a simple network configuration which was chosen in consultation with other researchers interested in assured forwarding. This is a simple configuration, which we believe, provides most insight in to the issues and on the other hand will be typical of a GEO satellite network.

We have addressed the following QoS issues in our simulation study:

- Three drop precedence (green, yellow, and red) help clearly distinguish between congestion sensitive and insensitive flows.
- The reserved bandwidth should not be overbooked, that is, the sum should be less than the bottleneck link capacity. If the network operates close to its capacity, three levels of drop precedence are redundant as there is not much excess bandwidth to be shared.
- The excess congestion sensitive (TCP) packets should be marked as yellow while the excess congestion insensitive (UDP) packets should be marked as red.
- The RED parameters have significant effect on the performance. The optimal setting of RED parameters is an area for further research.

## 2.1 Buffer Management Classifications

Buffer management techniques help identify which packets should be dropped when the queues exceed a certain threshold. It is possible to place packets in one queue or multiple queues depending upon their color or flow type. For the threshold, it is possible to keep a single threshold on packets in all queues or to keep multiple thresholds. Thus, the accounting (queues) could be single or multiple and the threshold could be single or multiple. These choices lead to four classes of buffer management techniques:

1. Single Accounting, Single Threshold (SAST)
2. Single Accounting, Multiple Threshold (SAMT)
3. Multiple Accounting, Single Threshold (MAST)
4. Multiple Accounting, Multiple Threshold (MAMT)

Random Early Discard (RED) is a well known and now commonly implemented packet drop policy. It has been shown that RED performs better and provides better fairness than the tail drop policy. In RED, the drop probability of a packet depends on the average queue length which is an exponential average of instantaneous queue length at the time of the packet's arrival [3]. The drop probability increases linearly from 0 to max_p as average queue length increases from min_th to max_th. With packets of multiple colors, one can calculate average queue length in many ways and have multiple sets of drop thresholds for packets of different colors. In general, with multiple colors, RED policy can be implemented as a variant of one of four general categories: SAST, SAMT, MAST, and MAMT.

Single Average Single Threshold RED has a single average queue length and same min_th and max_th thresholds for packets of all colors. Such a policy does not distinguish between packets of different colors and can also be called color blind RED. In Single Average Multiple Thresholds RED, average queue length is based on total number of packets in the queue irrespective of their color. However, packets of different colors have different drop thresholds. For example, if maximum queue size is 60 packets, the drop thresholds for green, yellow and red packets can be {40/60, 20/40, 0/10}. In these simulations, we used Single Average Multiple Thresholds RED.

In Multiple Average Single/Multiple Threshold RED, average queue length for packets of different colors is calculated differently. For example, average queue length for a color can be calculated using number of packets in the queue with same or better color [2]. In such a scheme, average queue length for green, yellow and red packets will be calculated using number of green, yellow + green, red + yellow + green packets in the queue respectively. Another possible scheme is where average queue length for a color is calculated using number of packets of that color in the queue [8]. In such a case, average queue length for green, yellow and red packets will be calculated using number of green, yellow and red packets in the queue respectively. Multiple Average Single Threshold RED will have same drop thresholds for packets of all colors whereas Multiple Average Multiple Threshold RED will have different drop thresholds for packets of different colors.

## 3. SIMULATION CONFIGURATION AND PARAMETERS

Figure 1 shows the network configuration for simulations. The configuration consists of customers 1 through 10 sending data over the link between Routers 1, 2 and using the same AF traffic class. Router 1 is located in a satellite ground station. Router 2 is located in a GEO satellite and Router 3 is located in destination ground station. Traffic is one-dimensional with only ACKs coming back from the other side. Customers 1 through 9 carry an aggregated traffic coming from 5 Reno TCP sources each. Customer 10 gets its traffic from a single UDP source sending data at a rate of 1.28 Mbps. Common configuration parameters are detailed in Tables 1 and 2. All TCP and UDP packets are marked green at the source before being 'recolored' by a traffic conditioner at the customer site. The traffic conditioner consists of two 'leaky' buckets (green and yellow) that mark packets according to their token generation rates (called reserved/green and yellow rate). In two-color simulations, yellow rate of all customers is set to zero. Thus, in two-color simulations, both UDP and TCP packets will be colored either green or red. In three-color simulations, customer 10 (the UDP customer) always has a yellow rate of 0. Thus, in three-color simulations, TCP packets coming from customers 1 through 9 can be colored green, yellow or red and UDP packets coming from customer 10 will be colored green or red. All the traffic coming to Router 1 passes through a Random Early Drop (RED) queue. The RED policy implemented at Router 1 can be classified as Single Average Multiple Threshold RED as explained in Section 3.

We have used NS simulator version 2.1b4a [9] for these simulations. The code has been modified to implement the traffic conditioner and multi-color RED (RED_n).

### 3.1 Experimental Design

In this study, we performed full factorial simulations involving many factors, which are listen in Tables 3 and 4 for two-color simulations and in Tables 5, 6 for three-color simulations:

- *Green Traffic Rates:* Green traffic rate is the token generation rate of green bucket in the traffic conditioner. We have experimented with green rates of 12.8, 25.6, 38.4 and 76.8 kbps per customer. These rates correspond to a total of 8.5%, 17.1%, 25.6% and 51.2% of network capacity (1.5 Mbps). In order to understand the effect of green traffic rate, we also conduct simulations with green rates of 102.4, 128, 153.6 and 179.2 kbps for two-color cases. These rates correspond to 68.3%, 85.3%, 102.4% and 119.5% of network capacity respectively. In last two cases, we have oversubscribed the available network bandwidth. The Green rates used and the simulations sets are shown in Tables 3 and 5 for two and three-color simulations respectively.
- *Green Bucket Size:* 1, 2, 4, 8, 16 and 32 packets of 576 bytes each, shown in Tables 4 and 6.
- *Yellow Traffic Rate* (only for three-color simulations, Table 6): Yellow traffic rate is the token generation rate of yellow bucket in the traffic conditioner. We have experimented with yellow rates of 12.8 and 128 kbps per customer. These rates correspond to 7.7% and 77% of total capacity (1.5 Mbps) respectively. We used a high yellow rate of 128 kbps so that all excess (out of green rate) TCP packets are colored yellow and thus can be distinguished from excess UDP packets that are colored red.
- *Yellow Bucket Size* (only for three-color simulations, Table 6): 1, 2, 4, 8, 16, 32 packets of 576 bytes each.

- *Maximum Drop Probability:* Maximum drop probability values used in the simulations are listed in Tables 4 and 6.

- *Drop Thresholds* for red colored packets: The network resources allocated to red colored packets and hence the fairness results depend on the drop thresholds for red packets. We experimented with different values of drop thresholds for red colored packets so as to achieve close to best fairness possible. Drop thresholds for green packets have been fixed at {40,60} for both two and three-color simulations. For three-color simulations, yellow packet drop thresholds are {20,40}. Drop threshols are listed in Tables 4 and 6.

In these simulations, size of all queues is 60 packets of 576 bytes each. The queue weight used to calculate RED average queue length is 0.002. For easy reference, we have given an identification number to each simulation (Tables 3 and 5). The simulation results are analyzed using ANOVA techniques [6] briefly described in Section 8.

### 3.2 Performance Metrics

Simulation results have been evaluated based on utilization of reserved rates by the customers and the fairness achieved in allocation of excess bandwidth among different customers.

Utilization of reserved rate by a customer is measured as the ratio of green throughput of the customer and the reserved rate. Green throughput of a customer is determined by the number of green colored packets received at the traffic destination(s). Since in these simulations, the drop thresholds for green packets are kept very high in the RED queue at Router 1, chances of a green packet getting dropped are minimal and ideally green throughput of a customer should equal its reserved rate.

The fairness in allocation of excess bandwidth among n customers sharing a link can be computed using the following formula [6]:

$$Fairness \quad Index = \frac{\left(\sum x_i\right)^2}{n \times \sum \left(x_i^2\right)}$$

Where $x_i$ is the excess throughput of the ith customer. Excess throughput of a customer is determined by the number of yellow and red packets received at the traffic destination(s).

## 4. SIMULATION RESULTS

Simulation results of two and three-color simulations are shown in Figures 2 and 3, where a simulation is identified by its Simulation ID listed in Tables 3 and 5. Figures 2a and 2b show the fairness achieved in allocation of excess bandwidth among ten customers for each of the two and three-color simulations respectively. It is clear from figure 2a that fairness is not good in two-color simulations. With three colors, there is a wide variation in fairness results with best results being close to 1. Fairness is zero in some of the two-color simulations. In these simulations, total reserved traffic uses all the bandwidth and there is no excess bandwidth available to share. Also, there is a wide variation in reserved rate utilization by customers in two and three-color simulations.

Figure 3 shows the reserved rate utilization by TCP and UDP customers. For TCP customers shown in Figures 3a and 3c, we have plotted the average reserved rate utilization in each simulation. In some cases, reserved rate utilization is slightly more than one. This is because token buckets are initially full which results in all packets getting green color in the beginning. Figures 3b and 3d show that UDP customers have good reserved rate utilization in almost all cases. In contrast, TCP customers show a wide variation in reserved rate utilization.

In order to determine the influence of different simulation factors on the reserved rate utilization and fairness achieved in excess bandwidth distribution, we analyze simulation results statistically using Analysis of Variation (ANOVA) technique. Section 4.1 gives a brief introduction to ANOVA technique used in the analysis. In later sections, we present the results of statistical analysis of two and three-color simulations, in Sections 4.2 and 4.3.

### 4.1 Analysis Of Variation (ANOVA) Technique

The results of a simulation are affected by the values (or levels) of simulation factors (e.g. green rate) and the interactions between levels of different factors (e.g. green rate and green bucket size). The simulation factors and their levels used in this simulation study are listed in Tables 3, 4, 5 and 6. Analysis of Variation of simulation results is a statistical technique used to quantify these effects. In this section, we present a brief account of Analysis of Variation technique. More details can be found in [6].

Analysis of Variation involves calculating the Total Variation in simulation results around the Overall Mean and doing Allocation of Variation to contributing factors and their interactions. Following steps describe the calculations:

*1.* Calculate the *Overall Mean* of all the values.

*2.* Calculate the individual effect of each level *a* of factor *A*, called the *Main Effect* of *a*:

$\text{Main Effect}_a = \text{Mean}_a - \text{Overall Mean}$

where, $\text{Main Effect}_a$ is the main effect of level *a* of factor *A*, $\text{Mean}_a$ is the mean of all results with *a* as the value for factor *A*.

The main effects are calculated for each level of each factor.

*3.* Calculate the *First Order Interaction* between levels *a* and *b* of two factors *A* and *B* respectively for all such pairs:

$\text{Interaction}_{a,b} = \text{Mean}_{a,b} - (\text{Overall Mean} + \text{Main Effect}_a + \text{Main Effect}_b)$

where, $\text{Interaction}_{a,b}$ is the interaction between levels *a* and *b* of factors *A* and *B* respectively, $\text{Mean}_{a,b}$ is mean of all results with *a* and *b* as values for factors *A* and *B*, $\text{Main Effect}_a$ and $\text{Main Effect}_b$ are main effects of levels *a* and *b* respectively.

*4.* Calculate the *Total Variation* as shown below:

$\text{Total Variation} = \sum(\text{result}^2) - (\text{Num_Sims}) \times (\text{Overall Mean}^2)$

where, $\sum(\text{result}^2)$ is the sum of squares of all individual results and Num_Sims is total number of simulations.

5. The next step is the *Allocation of Variation* to individual main effects and first order interactions. To calculate the variation caused by a factor *A*, we take the sum of squares of the main effects of all levels of *A* and multiply this sum with the number of experiments conducted with each level of *A*. To calculate the variation caused by first order interaction between two factors *A* and *B*, we take the sum of squares of all the first-order interactions between levels of *A* and *B* and multiply this sum with the number of experiments conducted with each combination of levels of *A* and *B*. We calculate the allocation of variation for each factor and first order interaction between every pair of factors.

## 4.2 ANOVA Analysis for Reserved Rate Utilization

Table 7 shows the Allocation of Variation to contributing factors for reserved rate utilization. As shown in Figures 3b and 3d, reserved rate utilization of UDP customers is almost always good for both two and three-color simulations. However, in spite of very low probability of a green packet getting dropped in the network, TCP customers are not able to fully utilize their reserved rate in all cases. The little variation in reserved rate utilization for UDP customers is explained largely by bucket size. Large bucket size means that more packets will get green color in the beginning of the simulation when green bucket is full. Green rate and interaction between green rate and bucket size explain a substantial part of the variation. This is because the definition of rate utilization metric has reserved rate in denominator. Thus, the part of the utilization coming from initially full bucket gets more weight for low reserved rate than for high reserved rates. Also, in two-color simulations for reserved rates 153.6 kbps and 179.2 kbps, the network is oversubscribed and hence in some cases UDP customer has a reserved rate utilization lower than one. For TCP customers, green bucket size is the main factor in determining reserved rate utilization. TCP traffic, because of its bursty nature, is not able to fully utilize its reserved rate unless bucket size is sufficiently high. In our simulations, UDP customer sends data at a uniform rate of 1.28 Mbps and hence is able to fully utilize its reserved rate even when bucket size is low. However, TCP customers can have very poor utilization of reserved rate if bucket size is not sufficient. The minimum size of the leaky bucket required to fully utilize the token generation rate depends on the burstiness of the traffic.

### 4.3 ANOVA Analysis for Fairness

Fairness results shown in Figure 2a indicate that fairness in allocation of excess network bandwidth is very poor in two-color simulations. With two colors, excess traffic of TCP as well as UDP customers is marked red and hence is given same treatment in the network. Congestion sensitive TCP flows reduce their data rate in response to congestion created by UDP flow. However, UDP flow keeps on sending data at the same rate as before. Thus, UDP flow gets most of the excess bandwidth and the fairness is poor. In three-color simulations, fairness results vary widely with fairness being good in many cases. Table 8 shows the important factors influencing fairness in three-color simulations as determined by ANOVA analysis. Yellow rate is the most important factor in determining fairness in three-color simulations. With three colors, excess TCP traffic can be colored yellow and thus distinguished from excess UDP traffic, which is colored red. Network can protect congestion sensitive TCP traffic from congestion insensitive UDP traffic by giving better treatment to yellow packets than to red packets. Treatment given to yellow and red packets in the RED queues depends on RED parameters (drop thresholds and max drop probability values) for yellow and red packets. Fairness can be achieved by coloring excess TCP packets as yellow and setting the RED parameter values for packets of different colors correctly. In these simulations, we experiment with yellow rates of 12.8 kbps and 128 kbps. With a yellow rate of 12.8 kbps, only a fraction of excess TCP packets can be colored yellow at the traffic conditioner and thus resulting fairness in excess bandwidth distribution is not good. However with a yellow rate of 128 kbps, all excess TCP packets are colored yellow and good fairness is achieved with correct setting of RED parameters. Yellow bucket size also explains a substantial portion of variation in fairness results for three-color simulations. This is because bursty TCP traffic can fully utilize its yellow rate only if yellow bucket size is sufficiently high. The interaction between yellow rate and yellow bucket size for three-color fairness results is because of the fact that minimum size of the yellow bucket required for fully utilizing the yellow rate increases with yellow rate.

It is evident that three colors are required to enable TCP flows get a fair share of excess network resources. Excess TCP and UDP packets should be colored differently and network should treat them in such a manner so as to achieve fairness. Also, size of token buckets should be sufficiently high so that bursty TCP traffic can fully utilize the token generation rates.

## 5. CONCLUSIONS

One of the goals of deploying multiple drop precedence levels in an Assured Forwarding traffic class on a satellite network is to ensure that all customers achieve their reserved rate and a fair share of excess bandwidth. In this paper, we analyzed the impact of various factors affecting the performance of assured forwarding. The key conclusions are:

- The key performance parameter is the level of green (reserved) traffic. The combined reserved rate for all customers should be less than the network capacity. Network should be configured in such a manner so that in-profile traffic (colored green) does not suffer any packet loss and is successfully delivered to the destination.

- If the reserved traffic is overbooked, so that there is little excess capacity, two drop precedence give the same performance as three.

- The fair allocation of excess network bandwidth can be achieved only by giving different treatment to out-of-profile traffic of congestion sensitive and insensitive flows. The reason is that congestion sensitive flows reduce their data rate on detecting congestion however congestion insensitive flows keep on sending data as before. Thus, in order to prevent congestion insensitive flows from taking advantage of reduced data rate of congestion sensitive flows in case of congestion, excess congestion insensitive traffic should get much harsher treatment from the network than excess congestion sensitive traffic. Hence, it is important that excess congestion sensitive and insensitive traffic is colored differently so that network can distinguish between them. Clearly, three colors or levels of drop precedence are required for this purpose.

- Classifiers have to distinguish between TCP and UDP packets in order to meaningfully utilize the three drop precedence.

- RED parameters and implementations have significant impact on the performance. Further work is required for recommendations on proper setting of RED parameters.

## 6. REFERENCES

[1] S. Blake, D. Black, M. Carlson, E. Davies, Z. Wang, W. Weiss, An Architecture for Differentiated Services, RFC 2475, December 1998.

[2] D. Clark, W. Fang, Explicit Allocation of Best Effort Packet Delivery Service, IEEE/ACM Transactions on Networking, August 1998: Vol. 6, No.4, pp. 349-361.

[3] S. Floyd, V. Jacobson, Random Early Detection Gateways for Congestion Avoidance, IEEE/ACM Transactions on Networking, 1(4): 397-413, August 1993.

[4] J. Heinanen, F. Baker, W. Weiss, J. Wroclawski, Assured Forwarding PHB Group, RFC 2597, June 1999. [3] V. Jacobson, K. Nichols, K. Poduri, An Expedited Forwarding PHB, RFC 2598, June 1999.

[5] V. Jacobson, K. Nichols, K. Poduri, An Expedited Forwarding PHB, RFC 2598, June 1999.

[6] R. Jain, The Art of Computer Systems Performance Analysis: Techniques for Experimental Design, Simulation and Modeling, New York, John Wiley and Sons Inc., 1991.

[7] S. Kota, "Multimedia Satellite Networks: Issues and Challenges," Proc. SPIE International Symposium on Voice, Video, and Data Communications, Boston, Nov 1-5, 1998.

[8] N. Seddigh, B. Nandy, P. Pieda, Study of TCP and UDP Interactions for the AF PHB, Internet Draft - Work in Progress, draft-nsbnpp-diffserv-tcpudpaf-00.pdf, June 1999.

[9] NS Simulator, Version 2.1, Available from http://www-mash.cs.berkeley.edu/ns.

Table 1: General Configuration Parameters used in Simulation

| | |
|---|---|
| Simulation Time | 100 seconds |
| TCP Window | 64 packets |
| IP Packet Size | 576 bytes |
| UDP Rate | 1.28Mbps |
| Maximum queue size (for all queues) | 60 packets |

Table 2: Link Parameters used in Simulations

| | |
|---|---|
| Link between UDP/TCPs and Customers: | |
| Link Bandwidth | 10 Mbps |
| One way Delay | 1 microsecond |
| Drop Policy | DropTail |
| Link between Customers (Sinks) and Router 1 (Router 3): | |
| Link Bandwidth | 1.5 Mbps |
| One way Delay | 5 microseconds |
| Drop Policy | DropTail |
| Link between Router 1 and Router 2: | |
| Link Bandwidth | 1.5 Mbps |
| One way Delay | 125 milliseconds |
| Drop Policy From Router 1 to Router 2 | RED_n |
| Drop Policy From Router 2 to Router 1 | DropTail |
| Link between Router 2 and Router 3: | |
| Link Bandwidth | 1.5 Mbps |
| One way Delay | 125 milliseconds |
| Drop Policy | DropTail |

Table 3: Two-color Simulation Sets and their Green Rate

| Simulation ID | Green Rate [kbps] |
|---|---|
| 1-144 | 12.8 |
| 201-344 | 25.6 |
| 401-544 | 38.4 |
| 601-744 | 76.8 |
| 801-944 | 102.4 |
| 1001-1144 | 128 |
| 1201-1344 | 153.6 |
| 1401-1544 | 179.2 |

Table 4: Parameters, which combinations are used in each Set of two-color Simulations

| | |
|---|---|
| Max Drop Drop Probability<br>{Green, Red} | {0.1, 0.1}, {0.1, 0.5}, {0.1, 1}, {0.5, 1}, {0.5, 1}, {1, 1} |
| Drop Thresholds<br>{Green, Red} | {40/60, 0/10}, {40/60, 0/20}, {40/60, 0/5}, {40/60, 20/40} |
| Green Bucket<br>(in Packets) | 1, 2, 4, 8, 16, 32 |

Table 5: Three-color Simulation Sets and their Green Rate

| Simulation ID | Green Rate [kbps] |
|---|---|
| 1-720 | 12.8 |
| 1001-1720 | 25.6 |
| 2001-2720 | 38.4 |
| 3001-3720 | 76.8 |

Table 6: Parameters, which combinations are used in each Set of three-color Simulations

| | |
|---|---|
| Max Drop Drop Probability<br>{Green, Yellow, Red} | {0.1, 0.5, 1}, {0.1, 1, 1}, {0.5, 0.5, 1}, {0.5, 1, 1}, {1, 1, 1} |
| Drop Thresholds<br>{Green, Yellow, Red} | {40/60, 20/40, 0/10}, {40/60, 20/40, 0/20} |
| Yellow Rate<br>[kbps] | 12.8, 128 |
| Green bucket Size<br>(in packets) | 1, 2, 4, 8, 16, 32 |
| Yellow bucket Size<br>(in packets) | 1, 2, 4, 8, 16, 32 |

Table 7: Main Factors Influencing Reserved Rate Utilization Results

| Factor/Interaction | Allocation of Variation (in %age) | | | |
|---|---|---|---|---|
| | In two-color Simulations | | In three-color Simulations | |
| | TCP | UDP | TCP | UDP |
| Green Rate | 8.86% | 31.55% | 2.21% | 20.41% |
| Green Bucket Size | 86.22% | 42.29% | 95.25% | 62.45% |
| Green Rate - Green Bucket Size | 4.45% | 25.35% | 1.96% | 17.11% |

Table 8: Main Factors Influencing Fairness Results in three-color Simulations

| Factor/Interaction | Allocation of Variation (in %age) |
|---|---|
| Yellow Rate | 41.36 |
| Yellow Bucket Size | 28.96 |
| Interaction between Yellow Rate and Yellow Bucket Size | 26.49 |

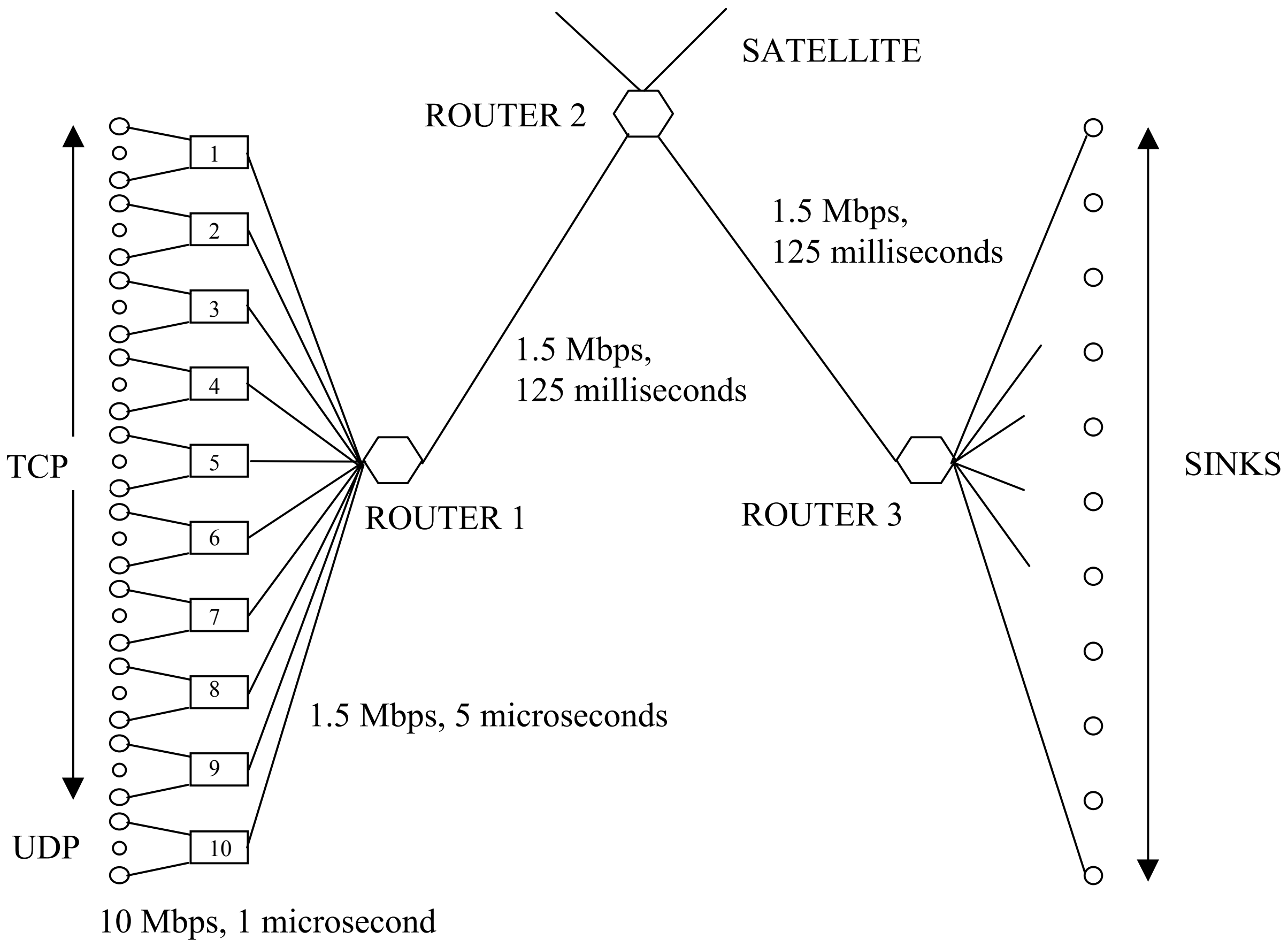

Figure 1. Simulation Configuration

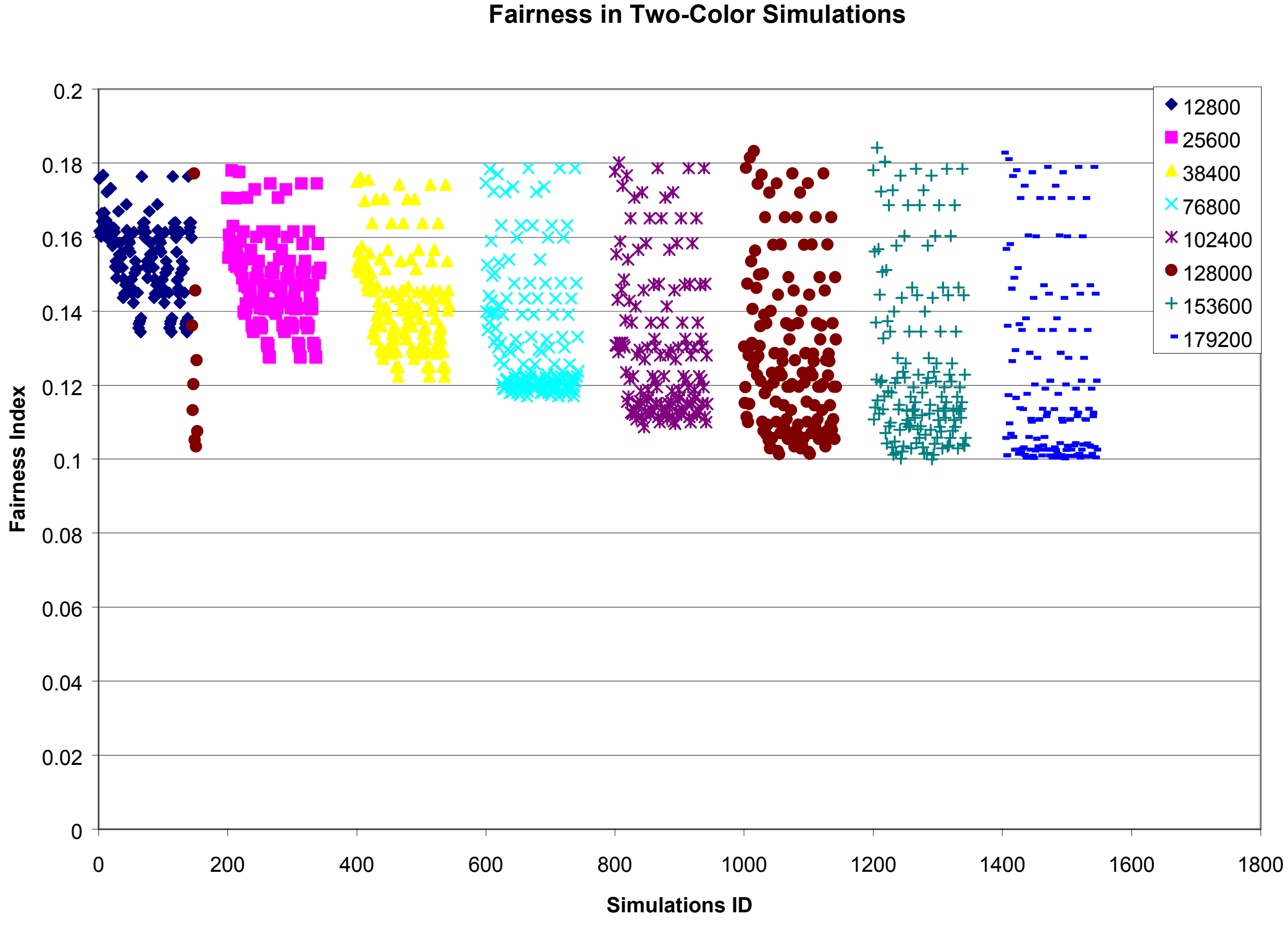

Figure 2a. Simulation Results: Fairness achieved in two-color Simulations with Different Reserved Rates

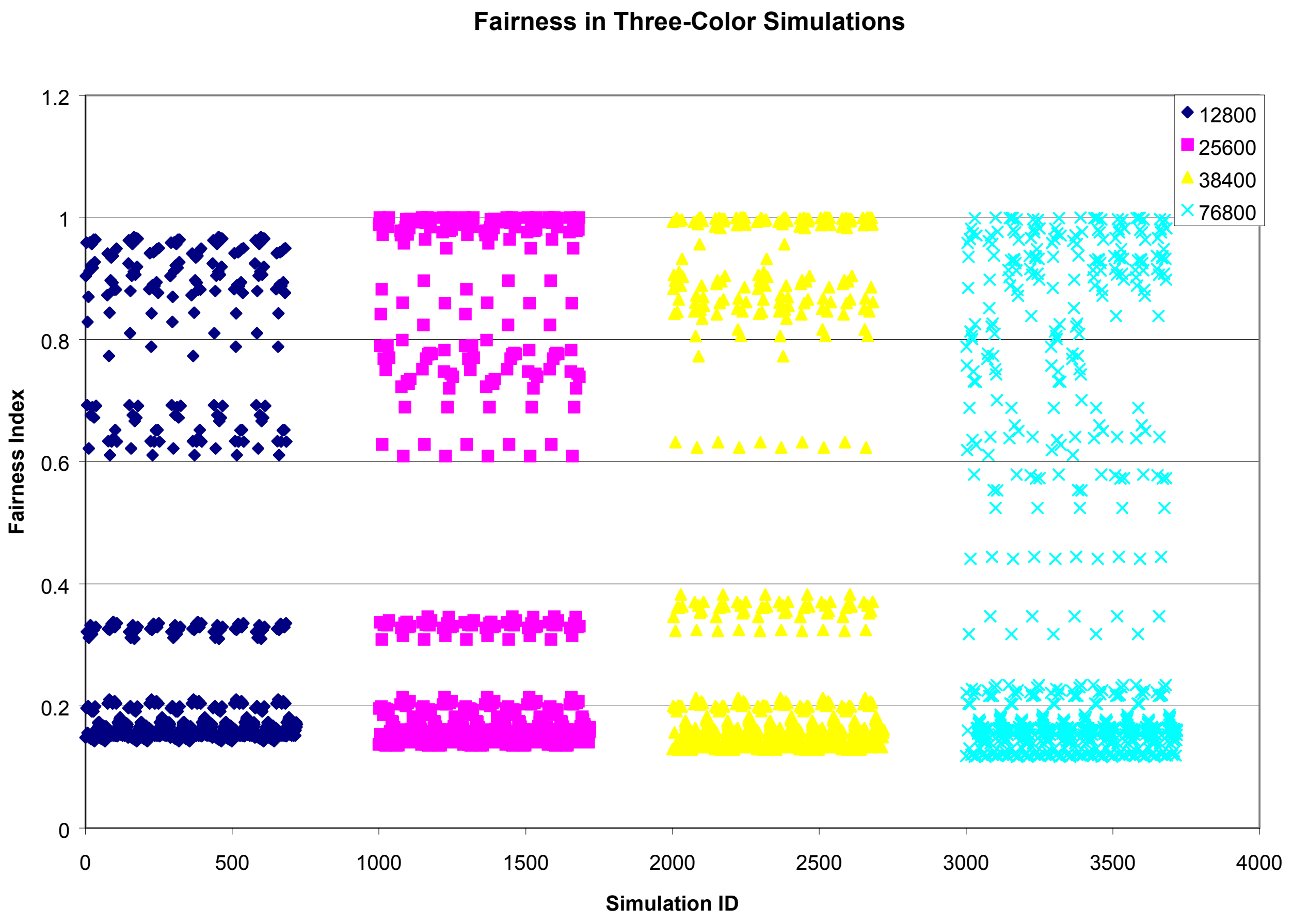

Figure 2b. Simulation Results: Fairness achieved in Three-Color Simulations with different Reserved Rates

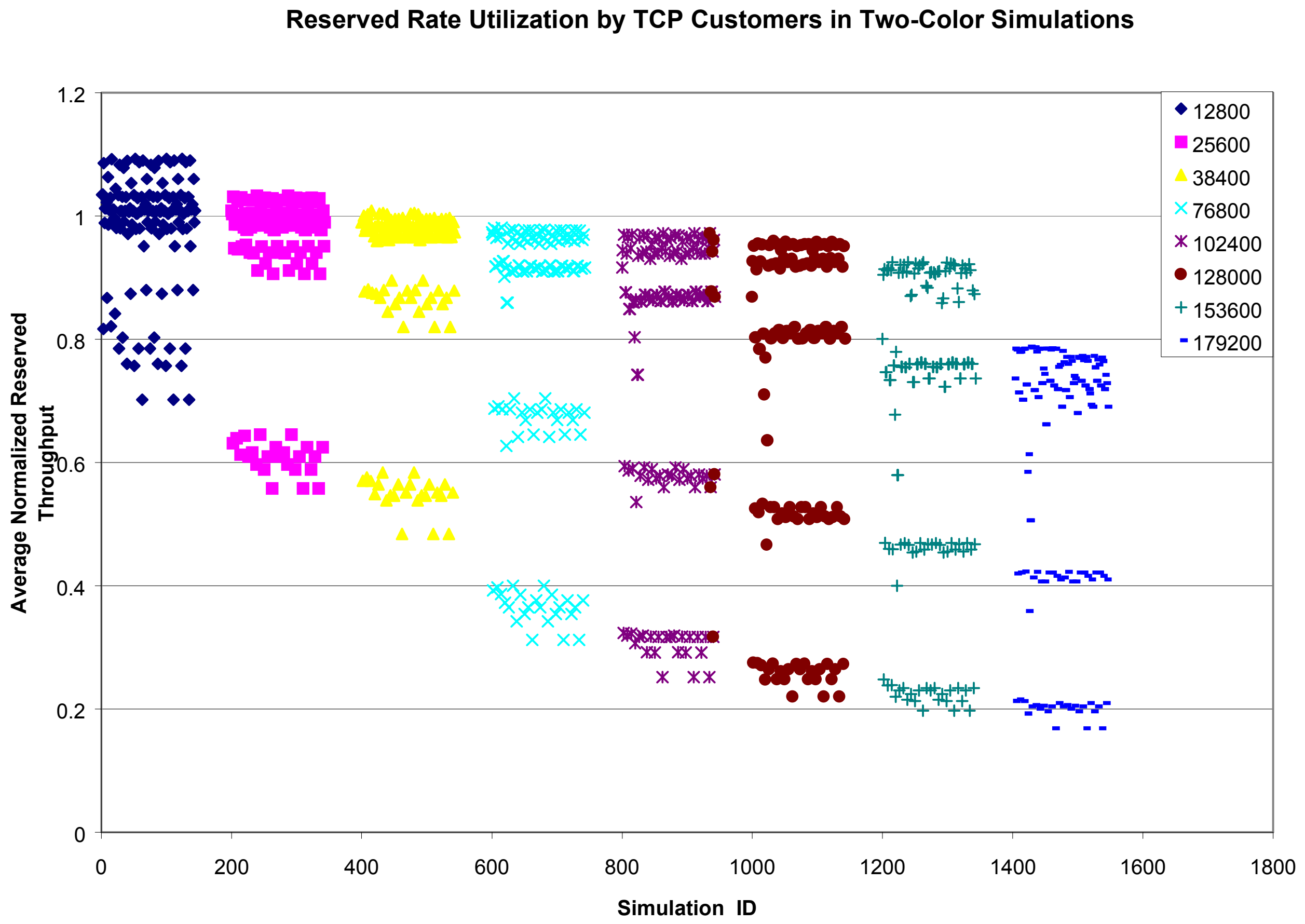

Figure 3a. Reserved Rate Utilization by TCP Customers in two-color Simulations

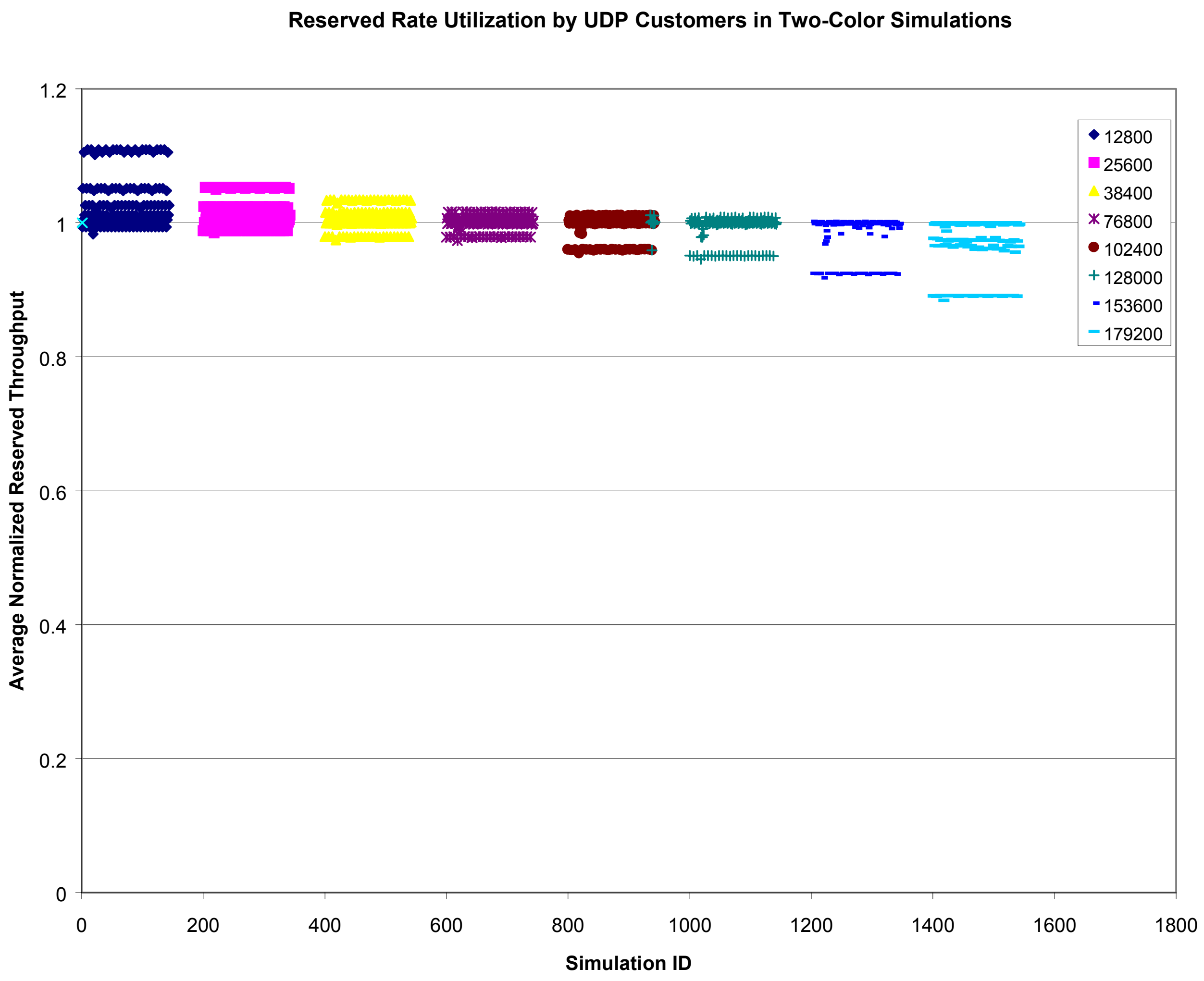

Figure 3b. Reserved Rate Utilization by UDP Customers in two-color Simulations

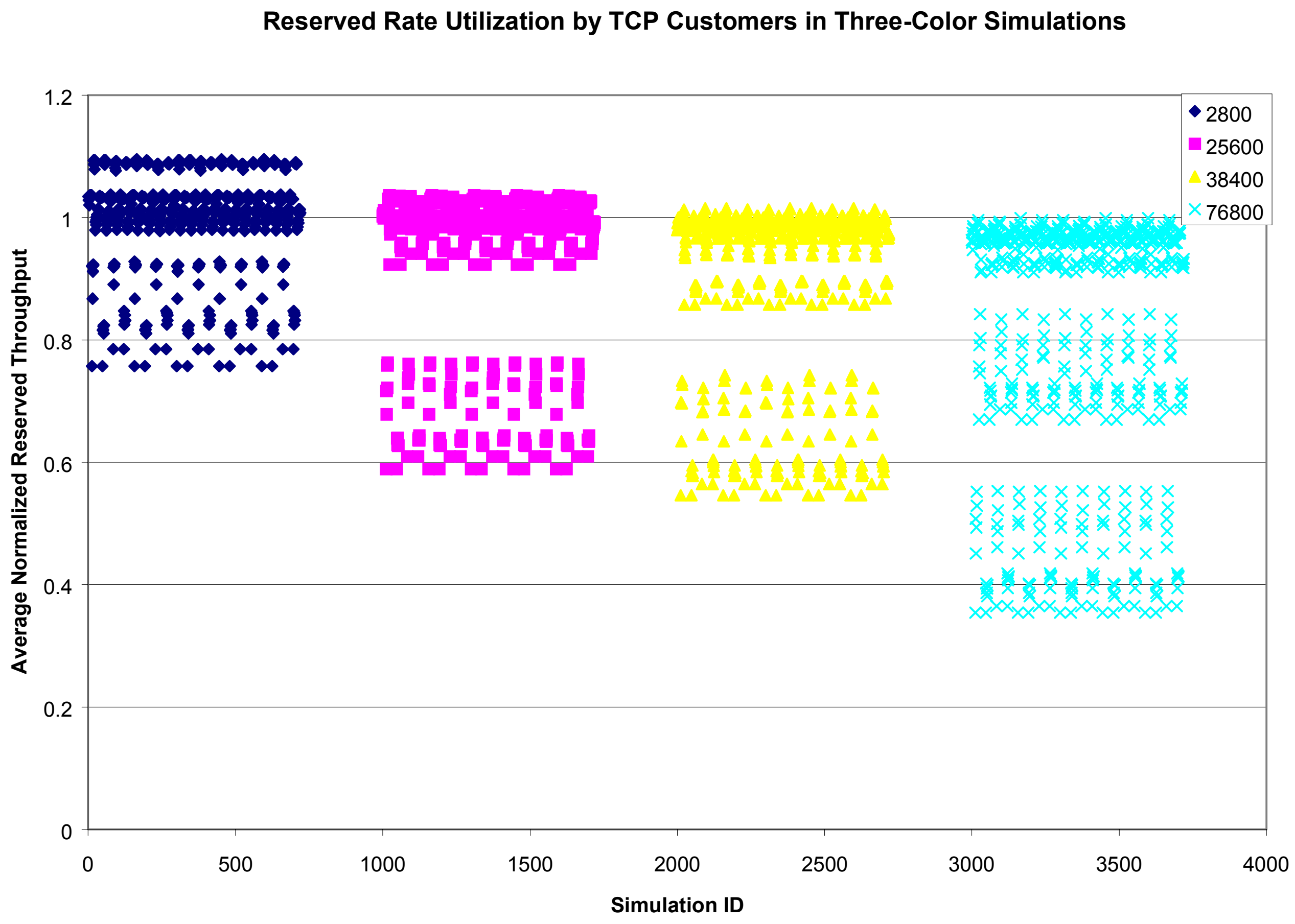

Figure 3c. Reserved Rate Utilization by TCP Customers in three-color Simulations

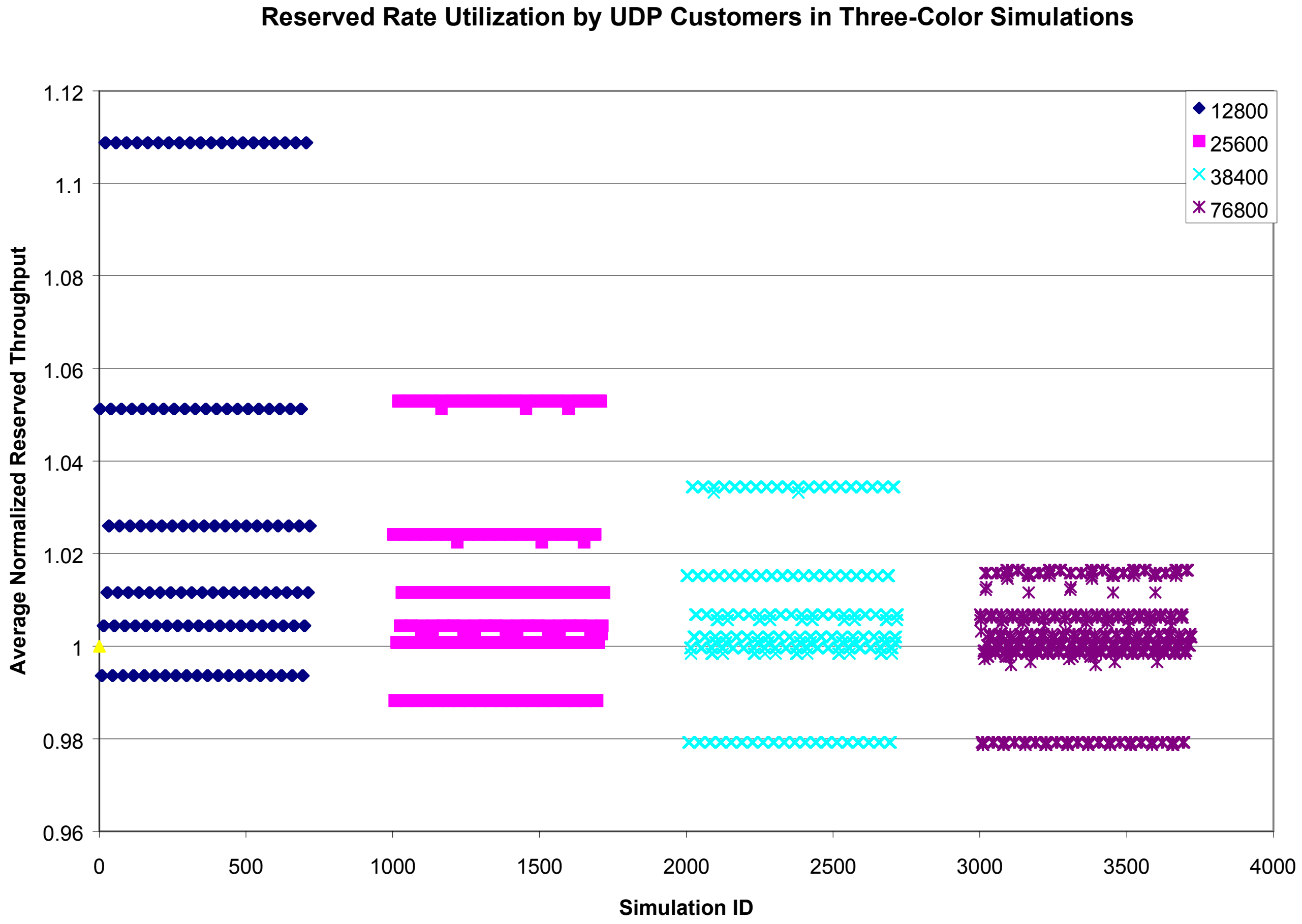

Figure 3d. Reserved Rate Utilization by UTP Customers in three-color Simulations